\documentclass[sigconf]{acmart}

\AtBeginDocument{%
  \providecommand\BibTeX{{%
    \normalfont B\kern-0.5em{\scshape i\kern-0.25em b}\kern-0.8em\TeX}}}

\copyrightyear{2024}
\acmYear{2024}
\setcopyright{rightsretained}
\acmConference[ASSETS '24]{The 26th International ACM SIGACCESS Conference on Computers and Accessibility}{October 27--30, 2024}{St. John's, NL, Canada}
\acmBooktitle{The 26th International ACM SIGACCESS Conference on Computers and Accessibility (ASSETS '24), October 27--30, 2024, St. John's, NL, Canada}
\acmDOI{10.1145/3663548.3675658}
\acmISBN{979-8-4007-0677-6/24/10}

\definecolor{brown}{rgb}{0.59, 0.29, 0.0}
\definecolor{darkgray}{rgb}{0.59, 0.59, 0.59}
\definecolor{tablegray}{gray}{.9}

\usepackage[normalem]{ulem}

\usepackage{pdfpages}

\begin{document}

\title{Context-Aware Image Descriptions for Web Accessibility}

\author{Ananya Gubbi Mohanbabu}
\affiliation{%
  \institution{The University of Texas at Austin}
  \city{Austin, TX}
  \country{USA}}
\email{ananyagm@utexas.edu}

\author{Amy Pavel}
\affiliation{%
  \institution{The University of Texas at Austin}
  \city{Austin, TX}
  \country{USA}}
\email{apavel@cs.utexas.edu}

\begin{abstract}

    Blind and low vision (BLV) internet users access images on the web via text descriptions. New vision-to-language models such as GPT-V, Gemini, and LLaVa can now provide detailed image descriptions on-demand. 
    While prior research and guidelines state that BLV audiences' information preferences depend on the context of the image, existing tools for accessing vision-to-language models provide only \textit{context-free} image descriptions by generating descriptions for the image alone without considering the surrounding webpage context.
    To explore how to integrate image context into image descriptions, we designed a Chrome Extension that automatically extracts webpage context to inform GPT-4V-generated image descriptions. 
    We gained feedback from 12 BLV participants in a user study comparing typical context-free image descriptions to context-aware image descriptions. We then further evaluated our context-informed image descriptions with a technical evaluation. 
    Our user evaluation demonstrates that BLV participants frequently prefer context-aware descriptions to context-free descriptions.
    BLV participants also rate context-aware descriptions significantly higher in quality, imaginability, relevance, and plausibility. 
    All participants shared that they wanted to use context-aware descriptions in the future and highlighted the potential for use in online shopping, social media, news, and personal interest blogs. 
\end{abstract}



\begin{teaserfigure}
  \includegraphics[width=\textwidth]{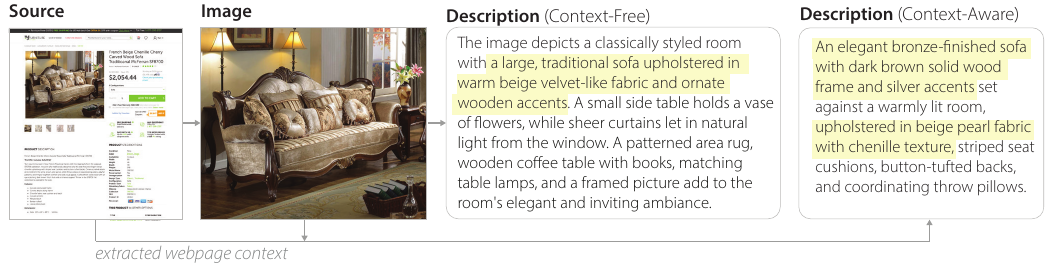}
  \caption{Our system provides context-aware descriptions for images by considering the image along with extracted details from the image source to craft a description that uses correct visual terminology (\textit{e.g.}, chenille texture rather than velvet) and focuses on the relevant item (\textit{e.g.}, the sofa rather than the room).}
  \Description{The figure shows information from left to right. The leftmost frame shows an e-commerce webpage where the image appears. An arrow points to the right from this picture to the image that the user selects on the webpage to obtain context-aware image description. An arrow from the selected image points to the right to the context-free description generated by GPT-4V. Arrows from the webpage labeled “extracted webpage context” and user selected image point to the rightmost frame that contains the context-aware description.}
  \label{fig:teaser}
\end{teaserfigure}

\maketitle

\section{Introduction}

Images are a primary way that people share information online. For example, people post selfies on social media, product photos on shopping sites, and scenic landscapes on travel websites. Blind and low vision (BLV) web users often use text descriptions to understand the content of images~\cite{wcag}, but images online often lack high-quality text alternatives such that the vast majority of images remain inaccessible~\cite{gleason2019ssocial,gleason2019making,gleason2019makingsocial,morris2016most,loiacono2009state,goodwin2011global}.
Recent vision-to-language models such as GPT-V~\cite{gpt_v}, Gemini~\cite{gemini}, and LLaVa~\cite{liu2023llava} can now convert images to detailed text descriptions on-demand, and new applications including an NVDA OpenAI plugin~\cite{nvdaopenai} and Be My AI~\cite{bemyai} provide access to such descriptions. 
Existing applications generate descriptions for the image alone, ignoring the surrounding context of the webpage where the image appears. 
In contrast, humans change what they describe in an image based on the image context~\cite{kreiss2022context}.
For example, a human description of the same room may be different for an AirBnB listing (\textit{e.g.}, number of seats, amenities, views) compared to a sofa listing (\textit{e.g.}, details of the sofa style).
Existing image description applications miss an opportunity to tailor their descriptions to specific webpage contexts. %

Prior work and image description guidelines have established that the context of the image, including where it appears, and its purpose, should inform the image description~\cite{stangl2020person,stangl2021going,diagramcenter_guidelines,w3c_tips,webaim_alttext}.
For example, Stangl et al. identified that the image source and information goal inform the information wants of BLV audience members~\cite{stangl2021going} and guidelines for image descriptions state that \textit{``context is key''}~\cite{diagramcenter_guidelines,webaim_alttext,w3c_tips} for deciding the content and terminology for image descriptions. 
Computer Vision researchers have explored using the surrounding context of an image to change the style of descriptions~\cite{gan2017stylenet,mathews2016senticap,shuster2019engaging} and improve the specificity of descriptions~\cite{everingham2006hello,biten2019good,srivatsan2023alt} (\textit{e.g.}, replacing named entities~\cite{biten2019good}, tailoring alt text to Twitter post text~\cite{srivatsan2023alt}).  
These prior approaches improve the description accuracy and specificity (\textit{e.g.}, by replacing ``woman'' with ``Mira''), but all prior approaches were created for an earlier class of vision-to-language models and thus were limited to simple context input (\textit{e.g.}, a short post text, a sentiment) and generated short descriptions. 
Prior work has not yet explored using rich webpage context to inform image descriptions or explored the impact of adding context for long descriptions provided.
Such prior work has also not yet collected feedback from BLV audience members on automatically generated context-informed descriptions. 
Motivated by a rich history of accessibility research~\cite{stangl2020person,stangl2021going,kreiss2022context} and guidance~\cite{diagramcenter_guidelines,webaim_alttext,w3c_tips} on the importance of context in descriptions, we build on such knowledge to explore how to automatically tailor image descriptions to their webpage context.

In this work, we present a system to automatically provide on-demand context-aware image descriptions for BLV users browsing the web. 
We prototype our system as a Chrome Extension that describes images on-demand as the user browses the web.
In particular, when a user selects an image to describe, the extension sends an image with the webpage to our pipeline. 
Our pipeline first extracts relevant webpage context, including the webpage text, title, and URL, along with the image and its alt text. Our pipeline scores all webpage text according to its relevance to the image (\textit{e.g.}, based on position and content similarity), then provides visually relevant details from the extracted context to GPT-V to produce a context-aware image description.   
The user receives context-aware short and long descriptions in our Google Chrome Extension such that they can flexibly access additional detail. 

We evaluated our system accuracy with a pipeline evaluation to assess risks of hallucinations and subjectivity for context-aware descriptions and conducted a user study with 12 BLV participants who frequently used AI description tools comparing context-aware and context-free descriptions. 
All participants wanted to use our system in the future and reported enthusiasm for context-aware descriptions. Participants rated context-aware descriptions significantly higher than context-free descriptions across all metrics we measured (quality, imaginability, relevance, and plausibility).
Participants reported the context-aware descriptions to be focused on more relevant details than context-free descriptions (especially for online shopping images) and appreciated the inclusion of concrete terminology from the surrounding text (especially for news images). 
Participants highlighted that the system provided relevant visual details tailored to the webpage audience and thus would be useful in areas of expertise (\textit{e.g.}, car details on a car blog).
While prior work established a need for context-aware descriptions~\cite{stangl2020person,stangl2021going,kreiss2022context}, participants in our study also highlighted risks of new potential automated descriptions (\textit{e.g.}, specific details may produce unwarranted trust, privacy concerns).

In summary, we contribute: 

\begin{itemize}
    \item An approach for automatically generating rich \textit{context-aware} image descriptions meeting an existing demand.
    \item A technical evaluation assessing the risks of context-aware descriptions. 
    \item A user study with 12 BLV participants comparing context-aware to context-free image descriptions.
\end{itemize}

\section{Related Work}
As we aim to create context-informed image descriptions for images encountered on the web, our work relates to prior work in describing images on the web and considering context in image descriptions.

\subsection{Describing Images on the Web}
Images are a primary medium for online communication across social media, blogs, tutorials, and more. 
BLV web users often access such images by reading text descriptions of the image content with screen readers or Braille displays. The Web Content Accessibility Guidelines (WCAG) thus request that creators \textit{``provide text alternatives for any non-text content''}~\cite{wcag} --- and text descriptions have been the standard since 1995~\cite{berners1995html}. Still, creators and platforms often fail to provide image descriptions such that most images online lack high-quality alternative text~\cite{gleason2019ssocial,gleason2019making,gleason2019makingsocial,morris2016most,loiacono2009state,goodwin2011global}. 

To make images accessible, prior research proposed human-powered~\cite{bigham2006webinsight,gleason2020twitter,morash2015guiding,von2006improving,bigham2010vizwiz,salisbury2017toward}, automated~\cite{bigham2006webinsight,gleason2020twitter,wu2017automatic,bemyai,google_chrome_image_descriptions,guinness2018caption,huh2022cocomix}, and hybrid approaches~\cite{bigham2006webinsight,gleason2020twitter,mack2021designing,singh2024figura11y} to craft descriptions for images. 
Existing approaches for automated and hybrid authored descriptions recognize the potential for inaccuracies of automated methods and mitigate this risk with human authoring. 
For example, WebInSight~\cite{bigham2006webinsight} and TwitterA11y~\cite{gleason2020twitter} generated descriptions but fell back to human-produced captions when automation failed, while Mack et al.~\cite{mack2021designing} and Singh et al.~\cite{singh2024figura11y} explored human-AI co-created alt text.
The recent progress of vision-to-language models has sparked a new era of image accessibility tools that can produce detailed and high-quality descriptions on-demand. BLV users can use tools like OpenAI's GPT-V~\cite{gpt_v} and Google's Gemini~\cite{gemini} directly to upload images and receive descriptions, or use screen reader plugins (\textit{e.g.}, an NVDA plugin~\cite{nvdaopenai}) to get in-place descriptions while browsing the web. 
While such tools now produce more accurate and detailed descriptions, they lack the context afforded to human image description authors (\textit{e.g.}, where and why the image appears in a website), and thus miss the opportunity to deliver context-relevant details. We explore how to integrate context into such descriptions.

As image descriptions are static, prior research has also explored alternative formats to enable gaining more information on demand by accessing a location in the image~\cite{morris2018rich,lee2022imageexplorer,nair2023imageassist}, gaining information through progressive detail or information type~\cite{huh2023genassist,morris2018rich}, selecting a description modality~\cite{gleason2020making,morris2018rich}, or by asking questions about the image~\cite{stangl2018browsewithme,bemyai,morris2018rich}. These interactive techniques let BLV audiences get information on demand but require extra effort to access additional information. We investigate how context can improve the information that BLV audiences initially receive.

\subsection{Why Context Matters for Descriptions}
An image description should share what the BLV reader needs to know to understand the page content. 
Accessibility guidelines offer high-level guidance on how to write high-quality image descriptions~\cite{perkins_how,perkins_creating,webaim_alttext,afb_accessibility,veroniiiica_redcarpet,veroniiiica_photojournalism,w3c_tips,diagramcenter_guidelines} by encouraging writers to be succinct~\cite{diagramcenter_guidelines,w3c_tips,webaim_alttext} and to consider the \textit{context} of the image \cite{diagramcenter_guidelines,w3c_tips,webaim_alttext} as the optimal content, tone, and terminology for an image description depend on the use of the image. For example, a photo of a woman crossing a street may be described as \textit{``Mari smiles while crossing a crosswalk in New York.''} on Mari's social media, or \textit{``A woman wears a short-sleeve chiffon knee-length sundress with a pastel pink and blue floral pattern.''} on a dress shopping website. 

As image description requirements vary based on context, prior research and guidelines explored information wants across different types of image contexts (\textit{e.g.}, for social media posts~\cite{bennett2018teenssocial,stangl2020person}, memes~\cite{gleason2019making}, red carpet looks~\cite{veroniiiica_redcarpet}, journalism~\cite{stangl2020person, veroniiiica_photojournalism}, and generating images~\cite{huh2023genassist}).
Prior studies specifically considered the role of context in BLV audience members information wants~\cite{stangl2020person,stangl2021going} and preferences~\cite{kreiss2022context} for image descriptions. Stangl et al. identified that information wants change for images encountered across different sources (\textit{e.g.}, wanted appearance details for dating profile images)~\cite{stangl2020person}, and even for the same image encountered in different contexts (\textit{e.g.}, wanted attributes of a bazaar for news but details of shirts in the bazaar for e-commerce)~\cite{stangl2021going}. 
Kreiss et al. further found that BLV audience members consider context relevance in their image description quality ratings~\cite{kreiss2022context}.
These studies confirmed that image context is crucial to determine what information to include in image descriptions, but vision-to-language model tools do not yet consider the context of the image when providing image descriptions.
Our work seeks to explore the potential of context-aware image descriptions by contributing a technical approach for creating them and conducting interviews with blind and low-vision users who compare automatically generated context-aware descriptions to context-free descriptions. 

\subsection{Augmenting Descriptions with Context}
Prior work also considered adding external text context beyond the image to inform the linguistic style~\cite{shuster2019engaging,gan2017stylenet,mathews2016senticap,chunseong2017attend} and content~\cite{everingham2006hello,biten2019good, srivatsan2023alt} of automated descriptions. 
For example, to guide linguistic style, Computer Vision researchers have explored providing captions based on positive or negative sentiment~\cite{mathews2016senticap}, personality type~\cite{shuster2019engaging}, style~\cite{gan2017stylenet} (\textit{e.g.}, humorous, romantic), and personal history~\cite{chunseong2017attend}.
Prior work has also investigated informing the content of descriptions by gaining information from the surrounding context~\cite{everingham2006hello,biten2019good,srivatsan2023alt}. 
Everingham et al. identified visual characters in movie frames based on movie scripts~\cite{everingham2006hello} and Biten et al. used named entities from news articles to identify out-of-vocabulary entities for news image captions~\cite{biten2019good} (\textit{e.g.}, replace ``woman'' with ``Mira'', or ``farm'' with ``Cherryville Lane''), but such work were not aimed at accessibility and thus did not motivate or evaluate how context-rich descriptions may impact blind users. Further, such work investigates changing the named entities in the description rather than the description focus itself.
Our work is most related to Srivatsan et al.'s work that creates alternative text for Twitter images considering the associated Tweet text~\cite{srivatsan2023alt}. While such prior work supports images with short plain text context (\textit{e.g.}, 240 character Tweet text~\cite{srivatsan2023alt}, sentiment labels~\cite{mathews2016senticap}), we support web images with long complex context (the entire webpage). Thus, we explore how to prioritize relevant context details, ignore irrelevant context details, and consider a variety of types of context (\textit{e.g.}, alt text, webpage title, surrounding text) when crafting descriptions. 
All prior work also used earlier vision to language models that produced short single sentence descriptions (\textit{e.g.}, BLIP-2~\cite{li2023blip}), such that our work is the first to investigate rich adaption of multi-sentence descriptions to context (\textit{e.g.}, adding rich details, introducing and defining terminology). We also uniquely seek feedback from BLV audience members who are frequent users of AI description tools to learn about the trade-offs of automatic context-tailored descriptions.

\section{Designing Context-Aware Image Descriptions}

To design a system to provide context-informed image descriptions for images encountered on webpages, we first examined what type of context appeared around images on webpages to determine what type of context a system may consider and then we reviewed image description guidelines and prior literature to inform our technical pipeline. 
We synthesize these activities as system design goals.

\begin{figure}
    \centering
    \includegraphics[width=3.33in]{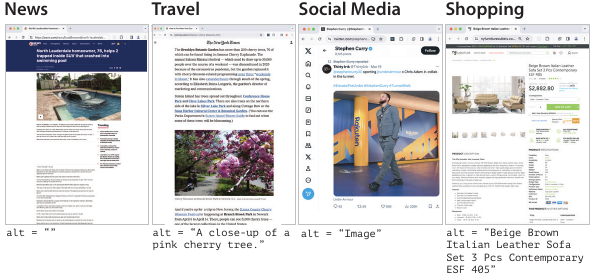}
    \caption{Webpage image context examples across news, travel, social media and shopping categories.}
    \Description{Four frames containing image contexts (screenshot of webpage and alt text of image in view) across news, travel, social media, and shopping categories from left to right. For the news article in the leftmost frame with an image of a car submerged in a pool, the alt text is missing. The frame to its right is a travel blog containing an image with pink cherry trees on either side of a road with alt text “A close-up of a pink cherry tree” written below the frame. The next frame is a screenshot of a post on X platform. It is an image of Stephen Curry in a gray suit. It contains the alt text “Image” below the frame. The last and rightmost frame is a screen grab of an e-commerce website selling sofas. It has the alt text “Beige Brown Italian Leather Sofa Set 3 Pcs Contemporary ESF 405” below the frame.}
    \label{fig:web-examples}
\end{figure}

\subsection{How Webpage Context Informs Descriptions}
Prior work identified that the context of the image at a high level (\textit{e.g.}, source, purpose) might inform image descriptions~\cite{stangl2020person,stangl2021going,kreiss2022context} and that for social media posts, the post text can inform image captions~\cite{gan2017stylenet,srivatsan2023alt}. However, the context on the web is complex (Figure~\ref{fig:web-examples}), and users may not want to manually enter information goals for each image. 
As high-level guidelines suggest that context is important for deciding what to describe~\cite{diagramcenter_guidelines,webaim_alttext,w3c_tips}, we aim to surface how low-level interpretations of context may impact image descriptions required for adding context to descriptions to inform system design. 
To broadly understand what type of context on the web we may extract to inform context-aware image descriptions, we examined webpages across a variety of contexts (examples in Figure~\ref{fig:web-examples}). We noted types of webpage context and potential impacts on descriptions (Table~\ref{tab:web-context}). 

Webpage context can \textbf{guide description focus} as the focus of the description depends on the purpose the image serves on the webpage~\cite{kreiss2022context}. For example, in the shopping example, the webpage title and adjacent text suggest the focus of the page is the 3 piece furniture set, and thus the furniture should be described in detail instead of the windows (Figure 2, Shopping).
Webpage context may also \textbf{guide description tone and terminology} as the tone and terms in the image description should be appropriate for the audience viewing the description~\cite{diagramcenter_guidelines}, the surrounding website text provides tone and terminology that may be used in the image description. For example, identifying Stephen Curry in the image may be useful for social media followers of Stephen Curry (Figure 2, Social Media). 
Terms in the article may help resolve potentially ambiguous elements in the image to \textbf{improve description accuracy}. For example, if the image displays a partially submerged vehicle the caption may help clarify that the vehicle is a Ford Explorer rather than a Honda CRV (Figure~\ref{fig:web-examples}, News). Finally, the image on the page alone and relative to other media and content may \textbf{guide level of detail and presence of the description}. For example, the purpose of a main image on an e-commerce page may be informative~\cite{wcag} and thus should receive detailed descriptions while small thumbnail images may be primarily for navigation and thus do not receive a detailed description~\cite{wcag} (\textit{e.g.}, Figure 2, Shopping). Purely decorative images based on the surrounding context may also not receive a description~\cite{wcag}. 

\begin{table}[]
\resizebox{3.33in}{!}{%
\begin{tabular}{@{}llll@{}}
\toprule
Category & Type & Examples & Function \\ \midrule
Content & URL & google.com & Purpose \\
 & Title & title tag & Purpose, visual concepts \\
 & Main Text & article, post text & Purpose, visual concepts \\
 & Tags & h1, h3, a, p & Text importance or purpose \\
 & Alt Text & alt tag & Image content, purpose, visual concepts \\
 & Caption & figcaption & Image content, purpose, visual concepts \\
 & Media & image, video & Purpose \\
Content Appearance & Size & width, height & Purpose, importance, relationship \\
 & Position & x, y, alignment & Purpose, importance, relationship \\
 & Color & color & Purpose, importance, relationship \\
 & Font & family, weight & Purpose, importance, relationship \\
 & Visibility & hidden & Purpose, importance, relationship \\
 & Other & texture, opacity & Purpose, importance, relationship \\
Image Appearance & Size & thumbnail, fullscreen & Purpose, importance, relationship \\
 & Position & top vs. mid article & Purpose, importance, relationship \\
 & Other & contrast, opacity & Purpose, importance, relationship \\ \bottomrule
\end{tabular}%
}
\caption{Examples of webpage context that may impact the visual interpretation of an image and how the image is described. Most of these webpage elements are selected intentionally by webpage authors (\textit{e.g.}, position of text content) to convey importance and structure to audience members, but others are dynamically added (\textit{e.g.}, advertisements).}
\label{tab:web-context}
\vspace{-10pt}
\end{table}

\subsection{Design Goals of Context-Aware Descriptions}
In crafting descriptions for BLV audience members we aimed to add context to descriptions while preserving existing image description guidelines. Based on prior literature and guidelines, we surface 5 key design goals: 
\begin{itemize}
    \item \textbf{D1.} Descriptions should be objective~\cite{diagramcenter_guidelines}.
    \item \textbf{D2.} Descriptions should be as concise as possible~\cite{w3c_tips,diagramcenter_guidelines}.
    \item \textbf{D3.} Prioritize information in descriptions to fit the context (\textit{e.g.,} main topic, purpose)~\cite{w3c_tips,stangl2021going}.
    \item \textbf{D4.} Language in descriptions should fit the context (\textit{e.g.,} names, places, objects)~\cite{w3c_tips,diagramcenter_guidelines}. 
    \item \textbf{D5.} The level of description provided should be informed by the context (\textit{e.g.}, decorative, informative)~\cite{wcag}. 
\end{itemize}
In our system, we address \textbf{D1-D4} but leave \textbf{D5} as an interesting avenue for future work. We select \textbf{D1-D4} as we intend to provide descriptions on-demand rather than for all images to save API costs. As users query for descriptions, we will assume that they would like enough detail to understand what is in the image. When context-aware descriptions become platform-supported, we may extend this work to \textbf{D5} to run our pipeline on less important decorative images. For \textbf{D1}, we aim to provide objectivity and reduce hallucinations in our technical pipeline (\textit{e.g.}, by focusing descriptions on details with visual evidence). For \textbf{D2}, we aim to provide concise descriptions that are succinct rather than unnecessarily wordy (e.g., repetitive or redundant). We provide concise descriptions at two levels of detail (short and long) to reflect that users have different preferences for description lengths~\cite{MacLeod2017}.
For \textbf{D3} and \textbf{D4}, we prioritize discussing visual details that are important to the context using language appropriate for the target audience.

\section{Prototype System}

\begin{figure*}
    \centering
    \includegraphics[width=\textwidth]{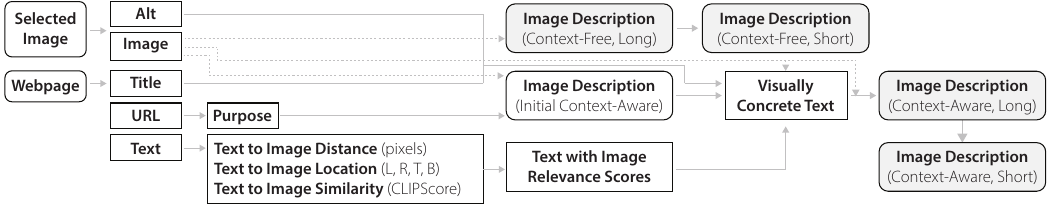}
    \caption{The system takes a webpage and a selected webpage image as selected by the user then provides context-aware descriptions based on both the webpage content and selected image.}
    \Description{The figure shows a diagram of system pipeline. From left to right, the diagram shows 6 groupings or columns, all connected by arrows pointing to the right. The first group includes Selected Image and Webpage. From Selected Image, there are two arrows pointing right to Alt and Image which are in the second group. From Webpage, there are three arrows pointing right to Title, URL, and Text (also in the second group). In the third column, there is an arrow pointing from URL to Purpose, and from Text there is an arrow pointing to the right to “Text to Image Distance (pixels), Text to Image Location (L, R, T,B), and Text to Image Similarity (CLIPScore). The Image also points to the right to Long Context-free Image Description and Initial Context-aware Image Description in the fourth group. Image, Alt, Title and Initial Context-aware Image Description all point to the right to Visually Concrete Text in the fifth column. The Long Context-free description points to the right to Short Context-free Image Description. The Visually Concrete Text and The Image point to the right to Long Context-Aware Image Description in the right in the sixth group. The Long Context-Aware Image Description points downwards to the Short Context-Aware Image Description.}
    \label{fig:system-diagraASSETSAAAm}
\end{figure*}

\begin{figure}
    \centering
    \includegraphics[width=3.33in]{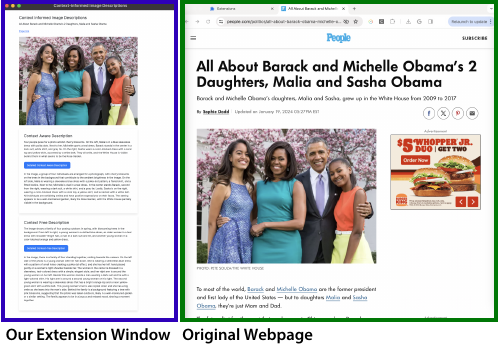}
    \caption{When a user clicks on an image in a website (right), our tool adds descriptions to the extension window (left).}
    \Description{An image of our chrome extension window on the left and the original webpage window with image on right. The extension window shows the user-selected image with short and long context-aware descriptions followed by short and long context-free descriptions. Users can optionally view the longer descriptions by clicking on the button below the shorter descriptions. The image selected is of Obama and his family.}
    \label{fig:interface}
\end{figure}

To gain feedback from blind and low vision users on automated context-aware descriptions, we built a prototype system to address guidelines \textbf{D1-D4}. Our prototype provides both context-free (existing) and context-informed (new) image descriptions at multiple levels of detail (Figure~\ref{fig:system-diagram}). To provide users access to descriptions, we created a Google Chrome Extension\footnote{https://github.com/UT-CS-HCI/context-aware-image-descriptions}
 that enables users to select an image as they browse the web to get context-aware descriptions for that image (Figure~\ref{fig:interface}). 

\subsection{Image Description Interface}
We created a Chrome Extension that users can install and activate for a webpage by clicking a button.
Our system opens a new extension window to provide descriptions.
While prior systems replaced alternative text directly in-place rather than in a separate window~\cite{bigham2006webinsight,gleason2020twitter}, our system preserves existing alternative text (Figure~\ref{fig:web-examples}) and enables flexible access to context-aware descriptions. 
Users can click an image to receive descriptions for that image in the extension window (Figure~\ref{fig:interface}). The extension window provides both short and long context-aware and context-free descriptions to mimic existing ``alt text'' and image descriptions (or ``long desc'') respectively. We provide the short description for both context-aware and context-free descriptions first, and users can optionally expand the longer descriptions on-demand. 
We include both context-aware and context-free descriptions to enable users to potentially recognize errors through transparency~\cite{huh2023genassist}.

\subsection{Image Description Pipeline}
When a user clicks an image, our system initiates a pipeline to provide context-aware descriptions (Figure~\ref{fig:system-diagram}). In particular, we extract relevant context from the webpage HTML including the webpage title, webpage URL, webpage text, and the alt text of the selected image. We then process this extracted context to distill information most likely related to the image and then we compose the final context-aware descriptions (short and long). 

\subsubsection{Extracting Webpage Context} 
As text on the webpage serves as the surrounding context for the image, we extract potentially relevant text from the webpage using the webpage's HTML. 
We extract the webpage \textbf{title} by using the \texttt{title} tag in the page as the title appears as the browser tab title and often communicates the key purpose of the webpage. We also extract the \textbf{URL} of the page and the webpage \textbf{text}. To extract the webpage text, we select all text elements with HTML tags typically used for text (\texttt{<a>}, \texttt{<p>}, \texttt{<span>}, and \texttt{<h1>} through \texttt{<h6>}).
We also extract the existing \textbf{image alt text} for the clicked image (\texttt{<alt>} tag) as the alternative text occasionally contains a useful short description of key image content that context-aware descriptions can further expand upon. 

\subsubsection{Processing Webpage Context} We initially attempted to add the raw HTML or extracted webpage context with the image directly to a vision language model (GPT-4V~\cite{gpt_v}) to craft a context-aware description. However, the model tended to repeat the webpage HTML or extracted text itself rather than describing the provided image. For example, GPT-4V summarized a news article rather than describing the provided image associated with the news article. Thus, to encourage the model to describe the image rather than the webpage context directly, we further processed the webpage context to surface potentially relevant details. 

First, we provided an \textbf{image relevance score} for each extracted text segment on the webpage that considers spatial and content relevance. Specifically, the image relevance score considers the relative position (proximity and layout) of each text segment to the image on the page and the content similarity between the image content and text segment content. We compute \textit{\textbf{proximity}} by first computing the distance from the center of the image to the closest edge of the text segment. We then normalize the proximity to achieve a score between 0-1 by dividing the computed proximity by the maximum possible proximity, i.e. the distance between the image and furthest text segment. We compute the \textit{\textbf{layout}} score by first providing a `top', `bottom', `left', or `right' property to describe the position of the closest edge of the text segment to the center of the image. We then assign a layout score of 0.8 for `top' and `bottom' or a layout score of 0.9 for `left' and `right'. 
We assign a higher score for left and right as the elements positioned as left and right positioning was less common overall (\textit{e.g.}, on e-commerce rather than blogs or news articles), but more likely to be used specifically for image details (\textit{e.g.}, e-commerce details to the left and right). 
The specific scores were intentionally close together such that they would primarily be used for breaking ties in proximity. 
We finally determined content \textit{\textbf{similarity}} using the CLIP score~\cite{hessel2021clipscore} between the text segment and image. We truncate text segments to 77 tokens to input them into CLIP to match CLIP's maximum input length. We immediately filter out all text elements with CLIP scores lower than 0.001 as such text segments are highly unlikely to be related to the image (\textit{e.g.}, often navigation bar links or advertisements) and use remaining text segments for the remainder of the pipeline. 
We compute the final image relevance score for the text segment as $imageRelevanceScore = 0.55*proximityScore + 0.1*layoutScore + 0.35*similarityScore$. We determined the score weights empirically in early testing to acknowledge the relative importance of position and content factors. 

To focus our descriptions on the website purpose, we also obtain a website \textbf{purpose} descriptor by extracting the URL of the webpage and prompting GPT-4 to \textit{``Identify
the domain of the web link, determine the category of the webpage in [ecommerce, news, educational...] and the purpose of the website in short.''} (see A.1 for the full prompt\footnote{We include select prompt details throughout this section for clarity, and include the full prompts with further instructions, formatting details, and sample outputs in the appendix. We refer to the relevant appendix section A.1-B.3 to indicate prompt location for each part of the pipeline.}). 
We then provided the image and initial website purpose to GPT-4V to obtain an \textbf{initial context-aware image description} with the prompt: \textit{``Describe the visual details of the element(s) in focus in the image for blind and low vision users to reinforce the purpose of the webpage''} (A.2). The initial context-aware description tailors the description of visual details from the image but lacks specific details from the surrounding webpage context, so we next extract relevant visually concrete text from the webpage.

\subsubsection{Extracting Visually Concrete Text.} 
To encourage our final description to attend to parts of the context that are relevant to the visual content in the image, we provide the initial context-aware image description, the alt text, the title, and the extracted context text with image relevance scores along with the image to GPT-4V to extract \textbf{visually concrete text} --- i.e. words or phrases from the initial image description (A.3) and context text segments (A.4) that can be clearly seen in the image --- and the elements in the image they are associated with.
For example, a visually concrete term in the context text may be \textit{``Rose Garden''} and the associated visible element in the image may be \textit{``Flowers in the background''}. 
To remove redundant visual concepts extracted from different parts of the context, we merge together visually concrete text (A.5) for similar visual concepts and again filter out visually concrete text (A.6) not present in the image to avoid hallucinations with GPT-4V.
We replace all names in the visually concrete text with a placeholders (\textit{e.g.}, person A, person B) using GPT-4V (A.7).

\subsubsection{Generating Context-Aware and Context-Free Descriptions} To generate context-aware descriptions we instruct GPT-4V to create a description based on the image and the visually concrete text we extracted (A.8). We include image relevance scores for the visually concrete text extracted from the website text to encourage the model to attend to website content that is more likely to be related to the image. We get a description back and replace all name placeholders with names using GPT-4V (A.9). As the model occasionally ignores the names, we also run this step 3 times then select a description by prompting GPT-4V to: \textit{``Choose the best description in [long context-aware descriptions] array
based on the highest number of visual details, named entities such as
names of people, location, objects, and objectivity''}. 
This process results in our final \textbf{long context-aware image description} and we also generate a description from the image alone to achieve a \textbf{long context-free image description} (B.1). For each description, we ask GPT-4V to make the descriptions more concise to obtain the \textbf{short context-aware description} (A.10) and \text{short context-free description} (B.3).

\subsubsection{Implementation} 
Our Chrome Extension interface was implemented using JavaScript. For the backend, we used a Python Flask server with a real-time Firebase~\cite{firebase2024} database to log context-aware and context-free descriptions for the user-selected images.  
To save on inference costs, we cache descriptions. Specifically, we use a Firebase real-time database to log the image, website, and description details. If the user revisits an image on a webpage, we retrieve a cached description directly from the database and display it.

\section{Pipeline Evaluation}
While we designed our pipeline to reflect our design guidelines, adding context to descriptions comes with the risk of adding hallucinations, subjective information, or irrelevant details into the descriptions. 
For example, our approach may capture a subjective description of a ``jaw-dropping house'' or a false claim of a ``basketball hoop'' from a vacation rental listing then add these details to the description. Our approach may also pick up details that are not relevant (\textit{e.g.}, a lamp is listed in suggested purchases, and it begins to describe the lamp in a couch listing). 
Thus, before we gathered user feedback, we first conducted a pipeline evaluation to assess the accuracy, objectivity, and relevance of our context-aware descriptions against several baseline descriptions.
Specifically, we evaluated short and long context-aware image descriptions from our pipeline and two baselines for accuracy, objectivity, relevance.

\subsection{Dataset \& Models}
We selected a set of 24 images from the web. These images were selected from four categories of websites: e-commerce, news, social media, and blogs. We selected e-commerce, news, and social media from prior work~\cite{stangl2020person} then selected blogs to generally cover other informational content (2 food, 2 travel, and 2 lifestyle). We selected 6 different websites from each of the four categories to represent a range of website structures, and selected one image from each website. 
The webpages were on average $12170$ characters long ($\sigma = 11487$) and had $1620$ words ($\sigma = 1592$ words).
We selected the websites and images such that the images had varying levels of alt text from no alt text to high-quality alt text. 
We ran our system on each image to generate a \textbf{context-free} long and short description (i.e. GPT-4V with no context) and a \textbf{context-aware} long and short description (i.e. GPT-4V with our pipeline to extract context). We also created a \textbf{context-HTML} baseline that provided GPT-4V the full HTML of the page as context as HTML would contain relevant semantic context details (\textit{e.g.}, title, alt text, web text) (i.e. GPT-4V with HTML as context). 
Across all methods, the long descriptions were around 2.5x the length of the short descriptions. Long descriptions were on average 172 words ($\sigma = 51$) for context-free, 136 words ($\sigma = 49$) for context-HTML, and 131 ($\sigma = 44$) for context-aware. Short descriptions were on average 56 words ($\sigma = 14$) for context-free, 49 words ($\sigma = 16$) for context-HTML, and 57 words ($\sigma = 20$) for context-aware. 
For each of the 24 images we evaluated 6 descriptions for a total of 144 descriptions (708 sentences in total).

\subsection{Analysis}
Two researchers, unaware of the description source, evaluated each sentence in across all descriptions for accuracy (accurate, inaccurate), and objectivity (objective, subjective).
\textbf{Accuracy} refers to whether or not each sentence contained a hallucination (inaccurate) or not (accurate) to assess whether our context-aware descriptions added hallucinations from the context (\textit{e.g.}, the descriptions mention that a person is wearing a hat, but no hat is present in the image). We considered a statement to be accurate if it contained zero errors and inaccurate if it had at least one error. An error is any text without matching visual evidence (e.g., ``4 shoes'' for 3 shoes, ``a hat'' for no hat). This strict binary measure is a lower bound on accuracy.
\textbf{Objectivity} assesses whether image descriptions contain subjective details without evidence in the picture (\textit{e.g.}, ``the cups are tastefully arranged on the table'' vs. ``the cups are on the table'') as image description guidelines suggest to be objective~\cite{webaim_alttext} to recognize if context-aware descriptions added subjective details. 
\textbf{Relevance} refers to the relevance of the sentence to the image given the context. For example, the color of a floorlamp in the background may be typically irrelevant on a dress shoppping website. While sighted people often provide image descriptions, BLV and sighted raters may disagree on relevance based on their lived experience, so we also assess relevance in a user evaluation with BLV participants.

Two researchers created the codebook by iteratively reviewing the data and refining codebook definitions. The researchers then both coded descriptions for a randomly sampled subset of the images (3 images with 99 total sentences, >10\% of the data) 
and achieved a moderate to substantial agreement across all codes (Cohen's $\kappa=0.53-0.79$). Then the researchers split the remaining descriptions to code independently. For the full codebook, see Supplementary Materials. 

Finally, we ran a named entity detector~\cite{spaCy-ent} across all descriptions to assess how often named entities (i.e. objects, people, locations, organizations that can be denoted with a proper name) were included in the descriptions.

\begin{figure}
    \centering
    \includegraphics[width=3.33in]{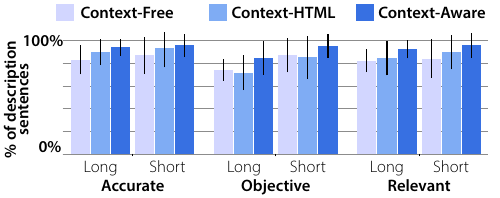}
    \caption{In the pipeline evaluation, we evaluated the accuracy, objectivity, and relevancy percentages of each description by coding if each sentence in the description did not contain a hallucination (accurate), did not contain subjective details (objective), and did not contain irrelevant details (relevant). Bars represent the average \% of accurate, objective, and description sentences for descriptions produced by the long and short version of each approach. The error bars are 95\% confidence intervals.}
    \Description{A vertical bar graph showing the percentage of description sentences for long and short versions of context-free, context-HTML, and context-aware descriptions across 3 criteria: accuracy, objectivity, and relevance. The y-axis is the percentage of description sentences. The x-axis is the accuracy, objectivity, and relevance criteria. Each criteria is evaluated for all sixes versions: long context-free, long context-HTML, long context-aware, short context-HTML, and short context-aware. The error bars are 95\% confidence intervals. The percentage of context-aware description sentences are higher for both long and short versions compared to the long and short versions of context-free context-HTML descriptions respectively for all 3 criteria.}
    \label{fig:pipeline-eval}
\end{figure}

\subsection{Results} Context-aware descriptions had a similar percentage of accurate, objective, and relevant sentences compared to the baseline context-free and context-HTML descriptions (Figure~\ref{fig:pipeline-eval}). Thus, adding context did not increase hallucinations, subjective statements, or irrelevant details. We also performed an error analysis of errors across all models to assess in what scenarios hallucinations occur.

Across all models, we saw several types of errors: plausible but not present visual objects (\textit{e.g.}, stating there is a factory near a parking lot of new cars, stating there is a small dog in a garden scene), plausible but inaccurate visual adjectives (\textit{e.g.} identifying a shiny pleated dress as striped due to lighting, describing a dark door as an open door), incorrect counts (\textit{e.g.}, a cabinet features 3 doors when it actually features 4 doors), and incorrect positioning (\textit{e.g.}, incorrectly stating a person is walking behind another person, stating the water level has reached a window when it is below the window). We specifically reviewed all named people extracted from the context with our context-aware descriptions to see if the entities had been misapplied by our name replacement step, but we did not find any errors in named entities. 

We further reviewed sentences that lacked visual evidence in the image (subjective sentences). Often the model provided statements without grounding in the image for both context-aware and context-free descriptions.  Common types of ungrounded statements included: adding subjective adjectives (\textit{e.g.}, calling a blue and red pattern ``harmonious''), providing subjective interpretation of the image as a whole (\textit{e.g.}, ``the overall impression is one of opulence and cultural heritige''), providing a guess about the image setting based on the colors (\textit{e.g.}, ``the golden hue suggests this image may have been taken during golden hour'') and adding explanations of image content not grounded in the image (\textit{e.g.}, ``typical of Indian bridal wear''). For context-aware images, we observed fewer subjective statements overall (our pipeline features multiple steps to assess visual grounding), but sometimes the subjective statement was uniquely derived from the context. 
For example, in an image of Billie Eilish on her Twitter with a post about her recent brand transformation and rise as a global phenomenon, the context-aware description includes the statement \textit{``The golden gleam of the awards [...] symbolizes both a brand transformation and her rise as a global phenomenon.''} whereas the context-free description features less specific interpretation \textit{``holding multiple gold gramophone trophies [...] conveying a sense of accomplishment and pride in their music industry achievements''}.

\begin{figure}
    \centering
    \includegraphics{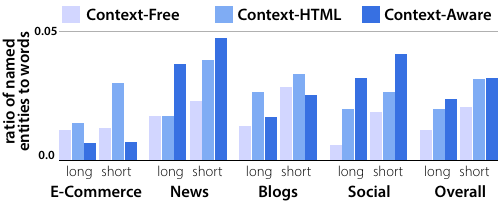}
     \caption{The ratio of named entities (\textit{e.g.}, proper nouns) to words for the short and long version of each description across all categories. Low means few named entities and high means many named entities.}
    \Description{A vertical bar graph showing the ratio of named entities to words for long and short versions of context-free, context-HTML, and context-aware descriptions across e-commerce, news, blogs, and social media categories. The y-axis is the percentage of description sentences. The x-axis is the categories e-commerce, news, blogs, social media and aggregate. Each criteria is evaluated for all sixes versions: long context-free, long context-HTML, long context-aware, short context-HTML, and short context-aware descriptions. The ratio of named entities to words are higher for both long and short versions of context-aware descriptions compared to the long and short versions of context-free and context-HTML descriptions for news, social media, and aggregate.}
    \label{fig:named-entities}
    \vspace{-20pt}
\end{figure}

Overall, our context-aware descriptions contain more named entities (\textit{e.g.}, proper nouns of people and places) than other methods (Figure~\ref{fig:named-entities}). 
Our approach excelled at including named entities for news and social media --- categories where specific names are often important. 
Our model selectively excluded named entities for e-commerce where specific names of people or places are typically not relevant. 
The blogs in our dataset consisted of educational content (2 food, 2 lifestyle, and 2 travel) such that named entities were important for some blogs (travel) but not others (food). 
Short descriptions obtained a higher number of named entities than long descriptions demonstrating the named entities were typically preserved during long to short description summarization for all description approaches.

\begin{table*}[]
\centering
\resizebox{\textwidth}{!}{%
\begin{tabular}{@{}lllllll@{}}
\toprule
PID & Age & Gender & Visual Impairment & Age of Onset      & Screen Reader(s)                      & Prior AI Tool Use                        \\ \midrule
1   & 26     & M   & Totally blind     & 16   & NVDA, Talkback                            & Google Chrome Descriptions                       \\
2   & 60     & F   & Totally blind     & Birth & Jaws                                      & Gemini, GPT                                      \\
3   & 29     & M   & Light perception  & Birth & NVDA, Jaws, TalkBack                      & Gemini, Seeing AI, Be My AI                      \\
4   & 32     & M   & Totally blind     & 7   & NVDA                                      & Gemini, Lookout, NVDA                            \\
5   & 52     & M   & Light perception  & 22   & Jaws                                      & Picture Smart AI                                 \\
6   & 55     & M   & Totally blind     & 7 & NVDA                                      & Picture Smart AI                                 \\
7   & 32     & M   & Totally blind     & 3   & NVDA, Narrator, Jaws, VoiceOver, TalkBack & Google Chrome Descriptions                       \\
8   & 29     & M   & Light perception  & Birth & VoiceOver                                 & Seeing AI, Be My AI                              \\
9   & 19     & M   & Light perception  & Birth & NVDA, VoiceOver                           & GPT, Claude, Co-pilot, Bing, Seeing AI, Be My AI \\
10  & 25     & F   & Totally blind     & 7   & Jaws, NVDA                                & Picture Smart AI                                 \\
11  & 33     & M   & Totally blind     & Birth & NVDA                                      & NVDA add on, OpenAI, Gemini, LLaMa               \\
12  & 23     & M   & Light perception  & Birth & VoiceOver, Jaws and NVDA                  & Seeing AI, Be My AI, Envision AI                 \\ \bottomrule
\end{tabular}%
}
\caption{Participant details for BLV participants in the user evaluation.} 
\label{tab:participants}
\end{table*}

\section{User Evaluation}
We then conducted a study with 12 blind and low vision AI description tool users to compare context-free to context-aware descriptions (on 6 images selected by us, and 2 images selected by participants) and provide open-ended feedback about benefits and risks of context-aware descriptions.

\subsection{Method}
Our user evaluation invited BLV participants to directly compare context-free and context-aware descriptions for 6 pre-selected images and 2 participant-selected images then provided an opportunity for open-ended feedback about risks and benefits of context-aware descriptions in a semi-structured interview. The study was an hour long, conducted in a 1:1 session via Zoom, and approved by our institution's IRB. We compensated participants with \$25 USD for their participation.

\begin{figure}
    \centering
    \includegraphics[width=3.33in]{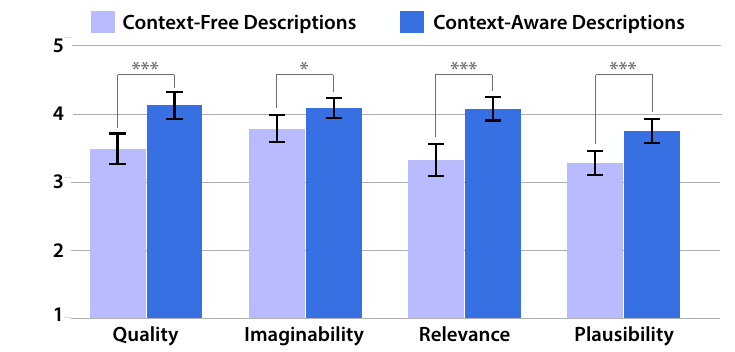}
    \caption{BLV participant ratings for quality, imaginability, relevance, and plausibility (on a scale from 1-low to 5-high) aggregated across all 6 images for Task 1. Asterisks indicate significance at the $p<0.001$ level (***), $p<0.01$ level (**), and $p<0.05$ level (*). We performed significance testing for all metrics with a Wilcoxon Signed Rank Test.}
    \Description{Image of vertical bar graph showing the BLV participants ratings for quality, imaginability, relevance, and plausibility for Task 1 of the User Study. The ratings are higher for context-aware descriptions compared to context-free image descriptions for all four criteria. Significance values for quality is p<0.001, imaginability is p<0.05, relevance is p<0.001, and plausibility is p<0.001.}
    \label{fig:task1-user-ratings}
\end{figure}

\subsubsection{Participants} We recruited 12 BLV participants who use screen readers to access information on the web and have experience using AI tools to describe images (Table~\ref{tab:participants}). We recruited participants using mailing lists. Participants used a variety of screen readers (NVDA, Talkback, Jaws, VoiceOver) and AI tools (\textit{e.g.}, Google Chrome Descriptions, Seeing AI, Be My AI, Picture Smart AI). Participants were totally blind (7 participants) or had some light perception (5 participants).

\subsubsection{Materials.} We selected 6 pre-selected website and image pairs for the study from our dataset collected in the pipeline evaluation to capture a variety of description types. We selected 3 images that contained people and 3 images that did not contain people across a range of categories (two images from both news and e-commerce and one image from blog and social media categories). For the 2 participant-selected images we invited participants to select their own image off a website of their choice. Participants chose images from blogs (tech, automobile, personal), news, educational articles, and more. For the 6 pre-selected images we used the descriptions obtained in the pipeline evaluation to keep the descriptions consistent across participants. For the participant-selected images, we installed our Google Chrome Extension on our computer to generate the image descriptions. See Appendix B for the full list of images and short descriptions for Task 1.

\subsubsection{Procedure} We first asked participants a series of demographic and background questions about their strategies and challenges with understanding image content while browsing the web. Participants then completed two tasks: a controlled pre-selected image task, and a open-ended participant-selected image task.

\textbf{Task 1}. In the first task, participants rated, selected, and provided open ended feedback on context-aware and context-free descriptions for 6 pre-selected images. For each image, we first invited the participant to browse the corresponding website and image for 2 minutes to understand the image context including the image alt text if present. Then, we provided one short description (either context-free or context-aware) with an option for users to extend the description to gain the long version of the description. We asked for participants to rate on a 5-point scale (from 1-low to 5-high) the quality, imaginability, relevance and plausibility of the description (evaluation measures selected from prior work~\cite{ kreiss2022context,stangl2021going}). We selected definitions of quality, imaginability, and relevance from prior work~\cite{kreiss2023contextref} (see Appendix E for metric questions).
We used plausibility~\cite{gardner2020determining} to assess the likelihood that a participant thinks a statement is true (similar to trust~\cite{MacLeod2017} and we use the terms interchangeably). We then provided the participant with the other description for the image (either context-aware or context-free) and repeated the rating questions. 
We randomized the sequence of images. We randomized and counterbalanced the order of context-aware and context-free descriptions for each participant and across participants for each image to mitigate ordering effects. We did not provide information about what type of description the participant was viewing during this stage to mitigate bias. 
After participants provided per-description ratings we asked participants to select what description they preferred (the first or second one they saw for the image) then provide open-ended feedback.

\textbf{Task 2}. In the second task, we mimicked real life use of the extension by inviting participants to provide an image from each of two websites of their choice. We provided participants context-aware and context-free descriptions for each image. We randomized the order to mitigate ordering effects. The descriptions were labelled (context-aware or context-free) to mimic real extension use. 
For each image, we initially provided users both context-aware and context-free short descriptions that they could optionally extend to see the longer version. We then asked participants to express preference between the two descriptions on a 5 point Likert Scale (from 1 - Strongly Prefer Context-Free to 5 - Strongly Prefer Context-Aware) then provide open-ended feedback about their preference. 

\textbf{Semi-Structured Interview}. We concluded a semi-structured interview asking participants about their perspectives on the benefits and drawbacks of context-aware descriptions and about potential future use of context-aware descriptions. 
See supplementary material for the full list of questions. 

\subsubsection{Analysis.} 
We recorded and transcribed the interviews. To examine participants' feedback on the context-aware image descriptions, one researcher read interview transcripts to derive themes through affinity mapping.

\begin{table}[t]
\resizebox{3.33in}{!}{%
\begin{tabular}{@{}lllllll@{}}
\toprule
 & \multicolumn{2}{l}{Context-Free} & \multicolumn{2}{l}{Context-Aware} &    &  \\ 
 & $\mu$ & $\sigma$ &$\mu$ & $\sigma$ & $p$ &  $Z$\\ \midrule
Overall Quality & 3.48 & 0.99 & 4.125 & 0.83 & < 0.001 & 4.17 \\
Relevance & 3.79 & 0.94 & 4.09 & 0.67 & < 0.05 & 2.56 \\
Imaginability & 3.33 & 1.08 & 4.08 & 0.8 & < 0.001 & 4.42 \\
Plausibility & 3.27 & 0.80 & 3.75 & 0.8 & < 0.001 & 3.58 \\ \bottomrule
\end{tabular}%
}
\caption{Mean ($\mu$) and standard deviation ($\sigma$) for all user ratings. Statistical tests performed with a Wilcoxon Signed Rank Test (p-value and Z displayed here).}
\label{tab:significance-testing}
\vspace{-20pt}
\end{table}

\begin{figure}
    \centering
    \includegraphics[width=3.33in]{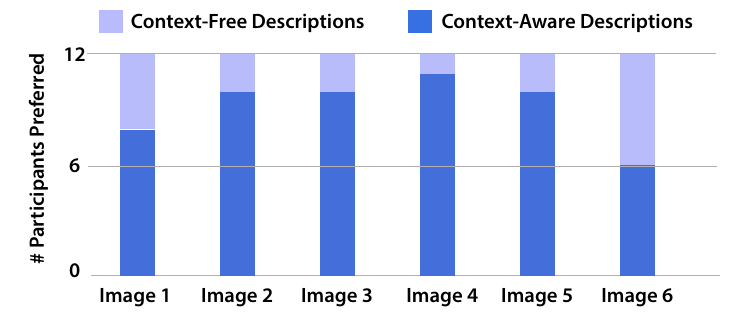}
    \caption{Preference between context-free and context-aware descriptions in Task 1.}
    \Description{The graph is a vertical bar that shows the number of participants that preferred the context-free vs. context-aware image description for the 6 pre-selected images in task 1. On the X axis is 1 to 6 images, and Y axis is the number of participants preferred. 8, 10, 10, 11, 10, and 6 of 12 participants preferred context-aware descriptions for images 1 to 6 respectively.}
    \label{fig:per-image-preference}
\end{figure}

\subsection{Results}
Overall, all participants reported they would want to use the Chrome Extension in the future to get context-aware descriptions about images on the web.
Participants preferred context-aware descriptions to context-free descriptions 76\% of the time in the first task (Figure~\ref{fig:task1-user-ratings}) and 67\% of the time in the second task (Figure~\ref{fig:task2-preference}). Participants in the first task also rated context-aware descriptions significantly higher than context-free descriptions (GPT-4V) across all metrics (quality, imaginability, relevance, and plausibility) (Figure~\ref{fig:task1-user-ratings}).

\subsubsection{Using Context Specific Terms for Visual Concepts} 
\begin{figure}
    \centering
    \includegraphics[width=3.33in]{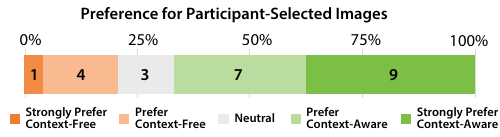}
    \caption{BLV participant description preferences for participant-selected images in Task 2. As each user selected 2 images each, there are 24 images in total. Participants preferred context-aware descriptions for the majority of images (16 of 24), but occassionally preferred context-free descriptions (5 of 24).}
    \Description{The image shows a horizontal stacked bar chart of description preferences for participant-selected images in Task 2 of the user study with strongly prefer context-free, prefer context-free, neutral, prefer context-aware, strongly prefer context-aware labels. Of the 24 images that were selected by the participants, context-free descriptions were strongly preferred and preferred for 1 and 4 images respectively, both context-free image description and context-aware image descriptions were equally preferred for 3 images, context-aware image descriptions were preferred and strongly preferred for 7 and 9 images respectively.}
    \label{fig:task2-preference}
\end{figure}

\begin{figure}
    \centering
    \includegraphics[width=3.33in]{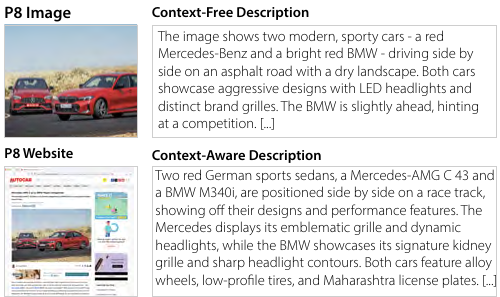}
    \caption{Example context-free and context-aware descriptions for a website and image selected by P8.}
    \Description{The image is from top to bottom. On the top-left side there is an image of two red cars on a tar road. To its right is the context-free description of the image that reads “The image shows modern sporty cars - a red Mercedes-Benz and a bright …on an asphalt road..The BMW is slightly ahead, hinting at a competition..”. To the bottom-left, is an image of the webpage where the image appears. The context-aware description of the image is present to the right of the image of the website and reads, “Two red German sports sedans, a Mercedes-AMG C43 … Mercedes displays its emblematic grille and dynamic headlights … and Maharashtra license plates…”.}
    \label{fig:car-example}
\end{figure}

All 12 participants stated that they found context-specific terms in context-aware descriptions to be useful for understanding the image. Participants frequently cited the context-specific terms in explaining why they chose a context-aware description over a context-free description (Figure~\ref{fig:per-image-preference} and \ref{fig:task2-preference}). 
While participant's existing visual description tools and context-free descriptions lacked context-specific terms, the context-aware descriptions often provided contextual details (\textit{e.g.}, ``Himalayas'' instead of ``Mountain Range'' in Image 6). 
Participants also highlighted that while their existing tools often left out names of people (\textit{e.g.}, GPT-4V provides \textit{``I'm sorry, but I can't identify...the people in the image.''} for images of people) our context-aware descriptions included the names of people (\textit{e.g.}, ``Billie Eilish'' instead of ``a woman'' in Image 1). 
P9 explained how context-aware descriptions improved imaginability: \textit{``I was immersed because its giving me the names, my brain took a second parsing when it said `a man', `woman' in the other description, but [the context-aware descriptions] gives me names and I could easily relate. It tells me about the White house, more specifics and details"}. 
P8 noted that the context-specific terms were particularly helpful when browsing an image he selected in the second task from his area of expertise: \textit{``I work for automotive industries, its got the key details very accurately, second one [context-aware description] is way way better.''} (Figure~\ref{fig:car-example}). 

While context-specific terms were beneficial when participants were familiar with the terms, participants noted difficulty interpreting terms they were not familiar with. P4 and P7 both mentioned that they did not know what ``Grammy'' trophies looked like so they noted they would prefer a combination of the context-aware description \textit{``multiple golden Grammy trophies against a backdrop of blurred Grammy trophies''} with the context-free description \textit{``multiple gold gramophone trophies''}. The long context-aware description included a similar explanation \textit{``multiple golden Grammy trophies, which are shaped like gramophones''} and future work may also explore enabling participants to define a term on demand~\cite{morris2018rich}. 
As we pre-selected 6 images in the first task, participants occasionally fell outside of the target audience of the page, such that the prevalence of context-specific terms could impede understanding. For example, P2 highlighted for a dress description that they appreciated that the context-aware description highlighted all the features of the dress (\textit{e.g.}, thin straps, hem) so that they could get a mental image, but these descriptions did not work well for P9 who mentioned: \textit{``I personally don't know what these mean, but I'm sure someone who shopped for dresses would like this description better.''}

\subsubsection{Including Relevant vs. Irrelevant Details} Participants rated context-aware descriptions as more relevant to the source than context-free descriptions (Table~\ref{tab:significance-testing}).
All participants highlighted that the context-aware descriptions were particularly strong at highlighting relevant details in the e-commerce images (Image 3 and Image 5). 
For example, in sofa listing (Image 5), P2 mentioned that while the context-free description explained the room, the context-aware description \textit{``focused on the sofa, that was what the image was about, I don't care bout the light coming in, table, or the rug! It went into details about the sofa, the wood color, back being tufted.''} While 10 of 12 participants preferred the context-aware description for the sofa due to the concentration on relevant details (ignoring irrelevant details), the remaining two participants appreciated that the context-free descriptions provided more information about the relevant context as it could also inform their buying decision, as P11 described: \textit{``it gave me some detail about the product but it gave more about context, sofa was able to accept the table, how tall it was and how it could match against other pieces of furniture''}. 
Beyond e-commerce, participants appreciated the focus on relevant over irrelevant details for topic-focused articles. For example, for an image on a food blog that P7 selected in the second task, the participant appreciated the focus on the food rather than the background: \textit{``It's much more descriptive, it describes the food first, which is ideal on the food blog, its accompaniments, and then the setting. Food is the main focus. It is also easy to imagine this way, seems more organized.''} 

Even when the context-aware and context-free descriptions included a similar distribution in the topics they described, the context-aware description often provided more specific details such that participants perceived the context-aware description as more relevant. For example, P10 selected a painting for a description and preferred the context-aware description as it provided a higher quanitity of relevant details (\textit{e.g.}, the context-aware description mentioned ``Wrinkles mark his forehead'' whereas the context-free description left this out, and highlighted a ``warm and indistinct'' background rather than just ``warm'') and included the subject's name (\textit{e.g.}, \textit{``an oil painting of Rabindranath Tagore''}).

\subsubsection{Subjective vs. Objective Details} 
Participants appreciated the level of objectivity in the details in the context-aware descriptions. Such objective details let participants form a mental image, as P9 reported: \textit{``It leaves no room for confusion, it tells me everything.''} (P9). However, users preferred context-free descriptions when context-aware descriptions omitted useful subjective details present in the context-free descriptions. For instance, for a news article on Prince Harry's ``protective'' gesture towards Meghan Markle (Task 1, Image 2), P3 and P4 expressed that they were able to imagine the image better with the context-free description because it provided hints about their relationship ``a man and a woman walk side by side, both with serious expressions''.
The context-aware description stated their names but lacked subjective details about their relationship or interaction. The action in the image was ambiguous (\textit{e.g.}, walking or standing).
P5 described that the context-aware description: 
\textit{``misses some of the details that they are walking side by side, which the [context-free] description tells us. So, both of them are okay. But if both of them were combined, then they could make a comprehensive description.}''.

\subsubsection{Trust} Participants found the context-aware descriptions to be more plausible compared to the context-free descriptions (Table~\ref{tab:significance-testing}). Context-aware descriptions described the images with terms from the image's context, which assured the participants that the descriptions are likely to be accurate. P7 said, \textit{``I trust it higher because some of the details it said on the webpage, it also said in the description.''}. Using the same terms as the website also made them less verbose unlike context-free descriptions which do not account for context while describing an image. P7, P9, and P10 also reported that for e-commerce websites, their trust in context-free descriptions was lower because context-free descriptions provided high-level overview of the product with focus on the background or details of the model showing the product, rather than product details.
In contrast, one participant expressed that they find it hard to trust detailed product descriptions: \textit{``Love the outfit description! but the more details it gets into, the less trust I have, because I automatically think that it is hallucinating and that's just my issues.''} (P7). P7 still rated the description generated by our system higher than the context-free description.
P5 also mentioned that the descriptions from both the context-free and context-aware models for products were surprisingly detailed and thus they would suspect potential hallucinations based on prior experience with AI models. 

\subsubsection{Future Use and Improvements} 
All participants they wanted descriptions to consider context in the future. 
All participants expressed their interest in using our extension in the future and stated that it would improve their general web browsing experience. 
Seven participants specifically expressed their interest in using context-aware descriptions for online shopping. Five participants reported that context-aware descriptions would be useful in seeing images of family and friends on social media to identify people, understand their expressions, and learn about their activities. 
Participants noted a range of other specific uses such as news articles with people, news articles about events, and automobile blogs. 
P12 mentioned that beyond web browsing, \textit{``It is really hard to find images for presentations that fit the context, I would use this tool for that too''}. 
4 participants wanted context-aware descriptions to also support chat, 4 participants mentioned that they wanted better support for text or graphical images (\textit{e.g.}, diagrams), and 2 participants mentioned that they wanted a central database to query for descriptions people already obtained (our extension supports caching, but this functionality could be extended). For text images and diagrams, our context-aware descriptions provided more details than context-free descriptions that only provided a high level overview. However, the details provided were not well-organized such that they were difficult to understand.

\subsubsection{Current Practice \& Comparison to Existing Approaches} 
Participants reported that they came across images on social media, blogs, news, CAPTCHAs and used them for programming, shopping, and analytics. Participants estimated that the images they come across online have alt text 30\% of the time on average (ranging from 5-40\% of time), and only a maximum of 20\% of those images had alt text that adequately described the image. 
While all participants frequently used AI tools for obtaining image descriptions, some participants mentioned that obtaining descriptions for online images was time-consuming with their current approach. 
Current applications required participants to download the images first they had to download the images first --- a step that is tedious and occasionally impossible with a screen reader. Participants found automated descriptions helpful, but for complex images participants reported that current AI generated descriptions contained hallucinations and often required the assistance of sighted people to verify description accuracy and obtain minor details. 

We asked participants how the descriptions across the study compared to their day to day descriptions and 8 participants reported the descriptions in the study were clearly better while P2, P3, P5 and P12 reported they were about the same overall, but that the context-aware descriptions were augmented with more useful details (\textit{e.g.}, names, places, object details). P2, P5, P8, P12 suggested that more descriptions were always better such that they wanted access to context-aware descriptions as a new technique.

\section{Discussion}
Our work contributed an approach for context-aware image descriptions on the web. Our work was motivated by an existing gap in tailoring image descriptions to their context as highlighted by the accessibility community and existing guidelines~\cite{stangl2021going,stangl2020person,diagramcenter_guidelines,webaim_alttext,w3c_tips}. 
Our technical pipeline represents the result of an iterative process to achieve relevant and context-aware descriptions with respect to this goal, contributing the first system to provide context-aware descriptions with modern vision to language models. 
The modern vision language models provide the capability for rich context input and multi sentence descriptions that can adapt to the webpage context in content and terminology use.
Our technical evaluation demonstrates the technical feasibility of our approach for achieving context-aware descriptions. Our study demonstrates that BLV participants who already use AI image description tools are excited about the potential of integrating context into their image descriptions. Our study also reveals benefits (\textit{e.g.}, a focus on relevant details, use of context-specific terms) and potential risks (\textit{e.g.}, increase plausibility may increase trust) of automated context-aware descriptions. 
We discuss several key trade-offs, limitations, and opportunities for future work.

\subsection{Context-Aware Descriptions for Expertise}
Current descriptions are one-size-fits-all by default, but description guidelines suggest tailoring not only content~\cite{stangl2021going} but also tone and terminology to the audience to match their knowledge and interests (\textit{e.g.}, college vs. grade school science class)~\cite{diagramcenter_guidelines,webaim_alttext,w3c_tips}. As our context-aware descriptions incorporate details from the surrounding context to shape the description, the descriptions follow the tone and terminology of the article. 
We originally intended to use the visual concepts from the context to improve specificity and accuracy (\textit{e.g.}, augment ``mountain range'' with ``Himilayas''), similar to Biten et al. improving vocabulary in short image captions~\cite{biten2019good}. 
However, the impact of using terminology that matched the context for long descriptions meant that for general audience news articles the approach often made the image easier to imagine. 
For websites with more specific audiences (\textit{e.g.}, a dress shopping store, a car blog) the descriptions became rich with context-specific terms such that the descriptions may be more enjoyable for experts to consume (\textit{e.g.}, the auto hobbyist reading an auto blog) and more challenging for novices to consume (\textit{e.g.}, a non-dress buyer reading about specific dress features). Over time, context-specific descriptions may have the potential to support furthering expertise in a domain of interest. 

Compared to our context-aware short descriptions that aim for conciseness, our context-aware long descriptions often included explanations of visual concepts. However, participants rarely accessed the long descriptions. 
To support ease of access for new visual terminology, we will explore combining context-free with context-aware descriptions such that a user may click on a term in the context-aware description to gain a context-free description of how the visual concept appears in the image. 
We will also take participant suggestions to integrate context-aware descriptions into chat such that participants can ask follow-up questions on demand.
Future work may also explore personalization to adapt the terminology use to prior description history or specific information goals of a browsing session~\cite{stangl2021going}.

\subsection{Risks and Trade-Offs of Context-Awareness}
Our prototype provides an opportunity to examine potential drawbacks of using context-aware descriptions in practice. 

\subsubsection{Trust and errors.} All vision to language models have some hallucinations, and our technical evaluation uncovered fewer hallucinations and subjective statements in context-aware descriptions compared to context-free descriptions.
Participants also rated the context-aware descriptions as significantly more plausible than context-free descriptions. 
A potential risk is that even though context-aware descriptions generate fewer errors, the errors may be more believable when they may come from the webpage context, such that the errors produced by context-aware descriptions may be more risky. 
We attempted to reduce context errors in the pipeline as much as possible by requesting that any extracted webpage context corresponded to image visual content. Thus, most context-related errors we observed were in subjective interpretation rather than complete hallucinations (\textit{e.g.}, stating an object that was not present).
Participants expressed mixed opinions on how details impacted their perceptions of plausibility. Most participants stated that more specific details made them trust the descriptions more, while other participants stated they trusted the detailed descriptions less (\textit{e.g.}, because the details went beyond their expected capabilities of AI models). Future deployments of context-aware descriptions may benefit from explanations of potential errors, and removing errors or making errors easier to detect by running the model multiple times so that the system or users could compare results~\cite{huh2023genassist}.

\subsubsection{Privacy. } Potential privacy concerns may also arise from adding context-aware descriptions to images. Participants expressed existing discomfort with needing to provide their images to an external server (\textit{e.g.}, OpenAI) to receive high-quality descriptions from new generation vision to language models. In our current implementation, the context offers an additional piece of information to an external server (\textit{e.g.}, the webpage where the image was viewed). While participants may not see a risk for mundane publicly accessible webpages, extending our tool to support image description in spaces with personal information in the context (\textit{e.g.}, in a family photo album) may raise additional concerns. Participants frequently expressed a preference for on-device models for both context-aware and context-free descriptions. As our system typically takes 30 seconds to 1 minute to produce a description and requires an OpenAI API key, one participant expressed that they wanted descriptions saved for future participants. We currently have this feature implemented, but in practice such a feature should be opt-in by people using context-aware descriptions. 

\subsubsection{Person identification. } Our system provides identification of people in an image when models (\textit{e.g.}, OpenAI, Gemini) explicitly prohibit identification of all people (\textit{e.g.}, for privacy), which participants noted disrupts their ability to imagine the image. Participants appreciated the ability of context-aware descriptions to identify people in articles across pre-selected images (common celebrities) and personally selected images (cricket players, Grace Hopper, a blogger). Context-aware descriptions may provide a unique opportunity to let BLV audience members access the identity of people in images without privacy risk as the only names that can be included in context-aware descriptions are already included in the context itself. We did not observe named entity errors in our dataset, such errors they may be present in images with more identified people (\textit{e.g.}, an image of a large group with many names in the context) due to vision language model capabilities and restrictions. 

\subsubsection{Limits to generalizability. } Our current pipeline works best for images that have relevant text context but the performance degrades in cases where the images lack text context (a photographer's page with only images) and for cases where the text content is unrelated or only loosely related to the images (a services page with stock photos). Our also system performs poorly also in cases where vision language models tend to perform poorly (\textit{e.g.}, structured image understanding~\cite{peng2024dreamstruct}). While our work created a proof of concept system and gained audience feedback on context-aware descriptions, future work can further explore limits to generalizability with larger scale evaluations and pipeline abalations. A fixed approach for context-aware descriptions is also not likely to fit user preferences across all users and contexts. For example, our system omitted subjective details (\textbf{D1}), but the diverse and context-specific user preferences about subjective details suggest the need for future work. We will explore how to let users customize the pipeline and prompts to set global and context-specific description preferences.

\subsection{Limitations and Future Work}
Our user study has several limitations. 
First, we recruited participants who were frequent users of AI image description tools to gain expert feedback. However, such users may have more knowledge about AI generated hallucinations than others. Second, the study was short-term so that we do not examine the impact of reading context-aware descriptions over time.
Long-term use may also allow users to learn context-relevant visual concepts or tire of visual details.
 Finally, participants did not use the Google Chrome Extension themselves to avoid installing the extension for a single short study. While the research team assured the interface was usable with a screen-reader, future work may further gain feedback on use of the extension in practice (similar to Gleason et al.~\cite{twitter-a11y}). 

Our technical approach and technical evaluation also offer some clear opportunities for extensions. First, we did not adapt the level of detail in our descriptions to fit the context, and we instead provided both long and short descriptions for every request. In the future, we will explore adapting the level of detail in the description according to other aspects of the context (\textit{e.g.}, size, visibility, complexity). 
Second, our technical evaluation used a strict binary measure to evaluate the accuracy of all statements in the pipeline evaluation. It may be valuable but non-trivial to evaluate major vs. minor errors as error severity may depend on its impact (e.g., importance of the detail, or likelihood of misleading the user). In addition, we require the user to query for descriptions from an image by clicking on it to save on API costs, but in the future we will explore providing descriptions for all images at once. In this scenario, future work may explore how to provide other images as context for the current image (\textit{e.g.}, if the image is in a set of three of a t-shirt, how does that change the descriptions?). Finally, our extension will offer more customization options in the future (\textit{e.g.} setting short or long descriptions as default).

\section{Conclusion}

We present a system to generate context-aware descriptions for images encountered on web using relevant context from the webpage HTML.
We evaluated the effectiveness of our system through a technical evaluation of description accuracy, objectivity, and relevance, and a user study with 12 BLV participants. In the user study, participants reported that the context-aware descriptions were more detailed, relevant, and plausible than context-free descriptions. All participants stated they wanted to use our system in the future. 
We aim to motivate future work to support people with disabilities performing everyday tasks on the web.

\section{Acknowledgements}

We thank Dr. Earl Huff Jr. from the School of Information at The University of Texas at Austin for his expert guidance during the initial stages of this research. His valuable insights informed our system design. We also thank our study participants for their time and valuable feedback on this work.

\bibliographystyle{ACM-Reference-Format}
\bibliography{sample-base}

\appendix

\section{System Pipeline Prompts}

We include prompts, input descriptions, and truncated sample outputs below, and include full sample outputs in supplemental material. 

\subsection{Webpage Purpose and Category}
We use the following prompt to determine the website purpose and category (categories are obtained from prior work~\cite{stangl2020person}): 
\textit{``Identify the domain of the web link, determine the category of the webpage in [ecommerce, news, educational, social media, entertainment, lifestyle, dating, job portals, or services] and the purpose of the website in short. Return the result only in a JSON format of '{"website": "name of the website", "category": "name of category", "purpose": "purpose of the website" }' with no additional text."} [Input: Webpage link]

\noindent \textbf{Example Output: } 
\begin{verbatim}
{
  "website": "people.com",
  "category": "entertainment",
  "purpose": "celebrity news and entertainment content"
}
\end{verbatim}

\subsection{Initial Context-Aware Image Description}
We use the following prompt and the website purpose to extract all the visual concepts of the elements in the image. \textit{``Describe the visual details of the element(s) in focus in the image for blind and low-vision users to reinforce the purpose of the webpage.''} [Input: Selected Image]

\noindent \textbf{Example Output: } The image shows four people standing in front of cherry blossoms in full bloom ... On the far left stands a young woman wearing a sleeveless dress with a blue base and adorned with a varying pattern of tiny dots ... modest neckline and a flared skirt, and she has her arm around another person next to her ...
To her right, a man stands with his arm comfortably around the young woman on his right ... wearing a classic dark suit with a light colored shirt and a dark tie ... His attire is formal, and he exhibits a polished look with his hair neatly trimmed ...

\subsection{Visually Concrete Text from Alt Text, Page Title, and Visual Description of the Image}
To extract the visually concrete texts and the elements in the image they refer to, we use the following prompt: \textit{``Identify all the visually concrete words and their attributes from the text. Verify if the visually concrete words can be associated with elements in the image. Return the result only in an array of JSON, in the format of [{vcw: "visually concrete word", element: "element associated with the visually concrete word"}] with no additional text such as starting with '''json'''. If no visually concrete words are present, return an empty JSON."} [Input: Image with Alt text, Webpage Title, Initial Context-Aware Description]

\noindent \textbf{Example Output: }Visual concrete text from visual description: 
\begin{verbatim}
[
  {
    "vcw": "people",
    "element": "four people standing"
  },
  {
    "vcw": "cherry blossoms",
    "element": "blossoms in full bloom in the background"
  },
  {
    "vcw": "smiling",
    "element": "expressions on people's faces"
  },
  {
    "vcw": "teal",
    "element": "color of the second woman's dress"
  },
  {
    "vcw": "watch",
    "element": "object on the second woman's left wrist"
  }, ...
]
\end{verbatim}

\noindent \textbf{Example Output: }Visual concrete text from alt text:
\begin{verbatim}
[
  {
    "vcw": "daughters",
    "element": "two younger females"
  },
  {
    "vcw": "family portrait",
    "element": "group photograph"
  },
  {
    "vcw": "Rose Garden",
    "element": "flowers in the background"
  },
  {
    "vcw": "Easter Sunday",
    "element": "not depicted"
  }, ...
]
\end{verbatim}

\noindent \textbf{Example Output: }Visual concrete text from page title: 
\begin{verbatim}
[]
\end{verbatim}
 
\subsection{Visually Concrete Text from Context}
To extract the visually concrete texts and the elements in the image they refer to, we use the following prompt: \textit{``Identify all the visually concrete words and their associated elements from the "text" field in the given JSON. If there are people/named entities present in the image, obtain their names from the highest "final\_score" in the JSON. Verify if the visually concrete words can be associated with elements in the image. The score of the visually concrete word is the "final\_score" field from which it is derived. Return the result only in JSON object in format of '[{vcw: "visually concrete word", element: "element associated with the visually concrete word", score: "final\_score"}]' with no additional text. If no visually concrete words are present, return an empty JSON.''} [Input: Image, Text from Webpage]

\noindent \textbf{Example Output: }
\begin{verbatim}
[
  {
    "vcw": "Barack",
    "element": "man in the middle with a tie",
    "score": 0.5320005792732359
  }, ...
]
\end{verbatim}

\subsection{Combining and Merging all Visually Concrete Text}
We combine and merge all the visually concrete text from alt text, webpage title, initial context-aware description, and the context. We retain the scores associated with elements from the visually concerete text from Context.
\textit{``Combine the visually concrete words that are associated with same elements, retain the score for the element if any entry for that element has a score. Keep all the named entities used to describe the elements. Return the result only in an array of JSON, with no additional text such as starting with '''json'''. If no similar elements are present, return the original JSON."} [Input: Image, JSON of Visually Concrete Texts from Alt Text, Webpage Title, Initial Context-Aware-Description, and Context]

\noindent \textbf{Example Output: }
\begin{verbatim}
[
  {
    "vcw": "Michelle",
    "element": "woman in the teal dress",
    "score": 0.17562086315826028
  },
  {
    "vcw": "dress",
    "element": "blue and polka-dotted dress on the left girl, 
    teal dress on the woman second from left, coral and 
    yellow dress on the right girl"
  },
  {
    "vcw": "White House",
    "element": "building partially visible in the background"
  }, ...
]
\end{verbatim}

\subsection{Filtering Visually Concrete Text}
In the combined JSON of visually concrete text, we filter the elements that are not visible in the image. \textit{``Generate a new JSON object from the given JSON by discarding entries whose "element" field is "none" or "not present". Return only the JSON with no additional text such as starting with '''json'''"} [Input: Image, Merged JSON of Visually Concrete Text]

\subsection{Replacing Name(s) of Person(s) with Placeholders}
We replace the names of the people (if present and if known) in the image using letters as placeholders. \textit{``If the names of person/people are known, only then assign {M, N, O, P...} (depending on the number of people in the image) to every person and return a JSON in the following structure: [{"placeholder": letter assigned to the name}, {"name": name of the person replaced}] with no additional texts. If there are no people, return an empty JSON."} [Input: Filtered JSON of Visually Concrete Text]

\noindent \textbf{Example Output: }
\begin{verbatim}
[
  {
    "placeholder": "M",
    "name": "Malia"
  }, ...
]
\end{verbatim}

\subsection{Long Context-Aware Image Description}
We generate a long context-aware description that is specific, detailed, relevant, and objective. \textit{``Describe the elements in focus in the image and their visual details for blind and low-vision users using all their visually concrete words (vcw) from the given JSON. If there is/are person/people in the image, refer to them in the description with the placeholder letters as given.  If there are no people in the image or their names are not present in the JSON, return the image description as is.`+ JSON.stringify(peopleVCW) +` Use the "scores" field to determine the priority of elements in the image to describe, higher score means higher priority to describe the element with its details. The goal is to make the image description specific and relevant. Return only the image description."} [Input: Image, JSON of Visually Concrete Text]

\noindent \textbf{Example Output: }
In the image, a group of four individuals ... with cherry blossoms on the trees in the background ... M is wearing a sleeveless blue dress with a polka-dot pattern ... N is clad in a teal dress... P is on the right, wearing a color-blocked dress with a coral top, a yellow skirt ... the Rose Garden, with the White House partially visible in the background.

\subsection{Adding Names to People in the Image (if available from Context)}
We replace the placeholders back with the names of the people in the image that we had retained initially. \textit{``If there is/are person/people in the image, replace the "placeholder" letters in the description with the corresponding "name" from the JSON. Ensure that the description is semantically and grammatically correct and return only the description. If there are no people in the image or their names are not present in the JSON, return the image description as is."} [Input: JSON of Names and their Placeholders]

\noindent \textbf{Example Output: }
In the image, a group of four individuals ... Malia is wearing a sleeveless blue dress with a polka-dot pattern ... Michelle is clad in a teal dress ... Sasha is on the right, wearing a color-blocked dress with a coral top, a yellow skirt ... the Rose Garden, with the White House partially visible in the background.

\subsection{Short Context-Aware Image Description}
\textit{``Refine the image description to make it more concise. If there is/are person/people in the image, replace the "placeholder" letters in the description with the corresponding "name" from the JSON. Ensure that the description is semantically and grammatically correct and return only the description.  If there are no people in the image or their names are not present in the JSON, return the image description as is."} [Input: Long Context Aware Description, JSON of Names and their Placeholders]

\noindent \textbf{Example Output: }Four people pose for a photo amidst cherry blossoms. On the left, Malia is in a blue sleeveless dress with polka dots. Next to her, Michelle sports a teal dress. Barack stands in the center in a dark suit, white shirt, and gray tie. On the right, Sasha wears a color-blocked dress with a coral top and yellow skirt, accented by a white belt. They all smile, and the White House is visible behind them in what seems to be the Rose Garden.

\subsection{Choosing the Best Long Context-Aware Description}
We choose the most objective, detailed, and context-aware description based on the number of named entities and visual details 
\textit{`Choose the best description in [long context-aware descriptions] array based on highest number of visual details, named entities such as names of people, location, objects, and objectivity. Return only the index number of the description once selected."} [Input: Array of Long Context Aware Descriptions]

\section{Baseline Prompts}
\subsection{Long Context Free Baseline Description}
We use the following prompt to obtain the long context-free baseline image description from GPT-4V, \textit{``Describe the image for blind and low-vision users."} [Input: Selected Image]

\subsection{Long Context HTML Baseline Description}
The following prompt is used to get the long context-HTML baseline description. \textit{``Describe the image for blind and low-vision users using the context."} [Input: Image, HTML of Webpage Text]

\subsection{Short Baselines}
We used the given prompt to obtain concise versions of long context-free and context-HTML baseline descriptions.
\textit{``Refine the image description to make it more concise."} [Input: Long Baseline]

\section{Images for User Study In Task 1}
\begin{table*}[h]
\centering
\resizebox{\textwidth}{!}{%
\begin{tabular}{@{}llp{.5\textwidth}p{.5\textwidth}l@{}}
\toprule
Image ID & Image & Image Alt Text & Webpage Title & Webpage Link \\ \midrule
1 & Grammy Awards & Image & No Conformity on X: "After Brand Transformation: With a brand transformation, Billie Eilish became a global phenomenon. She captivated audiences w/ her sound, aesthetic, \& authentic storytelling. Her brand evolution positioned her as an icon, earning her Grammy Awards. & \cite{x-billie}  \\
2 & Harry and Meghan & Prince Harry and Meghan Markle & Prince Harry's 'Protective' Gesture Over Meghan Markle Caught on Camera & \cite{meghan-harry}  \\
3 & Floral Dress & Color:Lilac Combo - Image 1 - Bluebell Floral Print V-Neck Sleeveless Maxi Dress & Free People Bluebell Floral Print V-Neck Sleeveless Maxi Dress & \cite{floral-dress}  \\
4 & Obama Family & Barack Obama, Michelle Obama, and daughters Malia (L) and Sasha (R) pose for a family portrait in the Rose Garden of the White House on Easter Sunday, April 5, 2015 in Washington, DC & All About Barack and Michelle Obama's 2 Daughters, Malia and Sasha Obama & \cite{obama-family}  \\
5 & Chenille Sofa & French Beige Chenille Cherry Carved Wood Sofa Traditioanal McFerran SF8700 & French Beige Chenille Cherry Carved Wood Sofa Traditioanal McFerran SF8700 – buy online on NY Furniture Outlet & \cite{sofa}  \\
6 & Annapurna Range, Himalayas & trekking in the Himalayas & Bucket List Travel: The Top 20 Places In The World & \cite{bucket-list}  \\ \bottomrule
\end{tabular}%
}
\caption{List of Pre-selected Images, their Alt Texts, and Webpage Titles used in Task 1 of User Study}
\label{tab:my-table}
\end{table*}

\section{Descriptions of Images in Task 1}

\begin{figure*}
    \centering
    \includegraphics[width=7in]{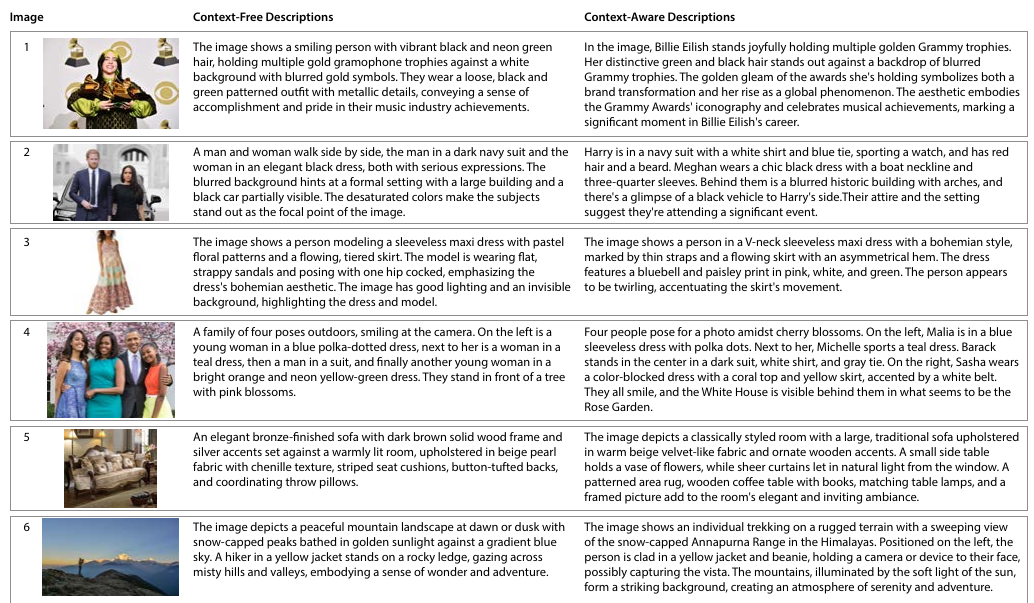}
    \caption{Example descriptions for Task 1 in the user study. See supplementary materials for full CSV of all descriptions.}
    \Description{The image contains the 6 pre-selected images and their short context-free and context-aware descriptions used in Task 1 of user study.}
    \label{fig:enter-label}
\end{figure*}

\section{User Study Evaluation Metrics Questions}

\begin{itemize}
    \item \textbf{Overall Quality}
How good is the description for overall nonvisual accessibility? 
    \item \textbf{Imaginability}
How well can you imagine this image in your mind?
    \item \textbf{Relevance} How well does the description capture the relevant aspects of the image?
    \item \textbf{Plausibility}
How much do you trust that the image description is correct?
\end{itemize}

\end{document}